\documentclass{vgtc}                          

\ifpdf
  \pdfoutput=1\relax                   
  \usepackage{graphicx}                
  \DeclareGraphicsExtensions{.pdf,.png,.jpg,.jpeg} 
\else
  \usepackage{graphicx}                
  \DeclareGraphicsExtensions{.eps}     
\fi%

\graphicspath{{figures/}{pictures/}{images/}{./}} 

\usepackage{caption}
\usepackage{subcaption}
\usepackage{microtype}                 
\PassOptionsToPackage{warn}{textcomp}  
\usepackage{textcomp}                  
\usepackage{mathptmx}                  
\usepackage{times}                     
\usepackage{cite}                      
\usepackage{tabu}                      
\usepackage{booktabs}                  
\usepackage{float}
\usepackage[table]{xcolor}
\usepackage{multirow}
\usepackage{siunitx}
\usepackage{todonotes}
\usepackage{setspace}

\usepackage{fancyhdr}
\fancypagestyle{firstpage}{
    \fancyhf{} 
    \fancyhead[L]{\textit{This is the author's preprint version of the article.}} 
}

\onlineid{0}

\vgtccategory{Research}

\preprinttext{To appear in an IEEE VGTC sponsored conference.}

\title{Evaluating Perceptual Deviations in Video See-Through Head-Mounted Displays while Utilizing Physical Touchscreens}

\author{Rudy De-Xin de Lange\thanks{rudy.de.lange@nlr.nl}\\ %
        \scriptsize Royal Netherlands Aerospace Centre %
\and Roemer Martin Bien Bakker\thanks{roemer.bakker@nlr.nl}\\
     \scriptsize Royal Netherlands Aerospace Centre %
\and Tanja Johanna Juliana Bos\thanks{tanja.bos@nlr.nl}\\ %
\scriptsize Royal Netherlands Aerospace Centre}
\vspace{-10em}
\begin{document}
\abstract{
Extended reality technology has become a useful tool in many applications, but still suffers from visual deviations that can hamper the utility of the technology. This paper discusses the types of persisting visual deviations experienced when observing the natural world through video see-through head-mounted displays. A generalizable method to measure the effect of these deviations on real-world interaction is designed and used in a human-in-the-loop experiment. The experiment compared video see-through sight through an head-mounted display with normal eyesight in a static set-up, focusing on (camera) lens distortions and display deviations. Participants interacted with a real touchscreen, locating the position of flashed markers shortly after disappearance comparing both conditions to check for deviations in position and time. Results show significant larger mean distance errors between the interaction locations and the original marker positions for video see-through compared to normal eyesight. Moreover, errors increase towards the screen periphery. No significant distance error improvement over time was found, however, response times did significantly decrease for both types of sight.}
\CCScatlist{
  \CCScatTwelve{Virtual Reality}{Augmented Reality}{Mixed Reality}{Head-Mounted Displays}
}







\maketitle
\thispagestyle{firstpage}

\vspace{0.5em}
\noindent 
{\small\setstretch{0}©2024 IEEE. Personal use of this material is permitted. Permission from IEEE must be obtained for all other uses, in any current or future media, including reprinting/republishing this material for advertising or promotional purposes, creating new collective works, resale, redistribution to servers or lists, or reuse of any copyrighted component of this work in other works. \textit{This is the author’s preprint version of an article accepted for publication at ISMAR 2024. The final version will be available on IEEE Xplore, and a DOI will be assigned upon publication. This preprint may be cited, but please refer to the final version for official citation.}}

\section{Introduction}\label{sec:introduction}
With the release of the Meta Quest 3 and the Apple Vision Pro, extended reality (XR) devices have gained traction within mainstream consumerism. XR, overarching virtual reality (VR), augmented reality (AR) and mixed reality (MR), has been researched for more than half a century. 
First functionally implemented in 1962, VR creates fully virtual environments (VEs) exclusively using digital content~\cite{Heilig1962Sensorama}.
Discussed in 1965~\cite{Sutherland1965UltimateDisplay}, with the first AR head-mounted display (HMD) introduced in 1968~\cite{Sutherland1968Jul}, AR enhances real-world experiences by overlaying digital content on the physical world. The Reality-Virtuality continuum defined in 1996 introduced the concept of MR, an in-between of VR and reality, spanning AR and augmented virtuality~\cite{Milgram1994Dec}, and is still used today~\cite{Skarbez2021Mar}.

Over the past decade, an increasing number of lower-cost XR head-mounted displays (HMDs) have emerged, increasing the attractiveness for individual, commercial, and research use~\cite{Castelvecchi2016,Cipresso2018Nov}. Commercial XR devices mostly target video-games, however other useful applications have been identified. Research has shown that XR can positively impact aircraft pilot training by reducing the amount of forgotten information during training~\cite{Macchiarella2004,Dymora2021Jan}. In turn, a shift within aircraft simulation and training is seen where, instead of large moving full flight simulators, more compact and lower-cost XR-based simulators are used~\cite{Oh2020Dec,Martins2021}. Often utilizing video see-through (VST) MR HMDs, these simulators combine VEs with real-world aspects, such as physical cockpit mock-ups. Even so, currently, only very limited training can be performed in these simulators~\cite{european_union_aviation_safety_agency_easa_fstd_2023}. Cybersickness, partly resulting from perceptual deviations in HMDs, impairs the achievable training duration, effectiveness and overall use of XR~\cite{stauffert_latency_2020}. Moreover, these deviations possibly affect real-world interaction performance and effectiveness~\cite{chang2020}. 

Mostly resulting from mismatches between the representation and perception of real versus virtual content, perceptual limitations were already identified in 1996~\cite{Drascic1996Apr} and still persist to this day, although in lesser degrees~\cite{Creem-Regehr2023}. Partly hard- and software dependent, perceptual issues can differ greatly between devices. How and which deviations affect proper usage of HMDs and affect interaction with VEs and the real-world is considered as the research gap. 

The aim of this paper is to identify and evaluate the effect of perceptual deviations in VST HMDs on real-world interaction. Section~\ref{sec:background} discusses perceptual deviations in XR and analyzes perceptual deviations of a Varjo XR-3 VST HMD as a use-case. Section~\ref{sec:experiment} establishes a method to measure the effect of perceptual deviations, subsequently used in a human-in-the-loop experiment. Section~\ref{sec:results} presents the experiment results, with possible future work and countermeasures based thereupon shortly discussed in Section~\ref{sec:discussion}. At last, Section~\ref{sec:conclusion} concludes this paper.

\section{Perceptual deviations in XR}\label{sec:background}
Contrary to VR, where fully enclosed HMDs are used to provide the user with only virtual content, MR combines virtual content with real-world aspects~\cite{Milgram1994Dec}. For this, two main types of MR HMDs exist, being VST and optical see-through (OST)~\cite{ballestin_registration_2021}. Typically, VST HMDs are, similar to VR HMDs, fully enclosed and utilize cameras to provide real-world information within an internal display. OST makes use of transparent surfaces that show the real-world as it is and on which digital content can be overlayed via collimation optics. OST technology is often used for MR applications where the focus is on the real-world with virtual enhancements. Contrary, VST is more common in applications where immersive VEs are combined with real-world aspects. 

\subsection{Video and optical see-through deviations}\label{subsub_perceptual_deviations_in_xr}
When perceiving MR, both VST and OST HMDs introduce perceptual deviations. The main deviations are discussed below.

\textit{Vergence-accomodation conflict (VAC)} -- arises from a fixed focal distance of a HMD of around 1.5~m. This causes a discrepancy in vergence and accomodation when objects that appear to be located on a different distance than this focal distance are visualized on the HMD~\cite{Kramida2016Jul,Zhou2021Dec}. Symptoms include eye-strain, fatigue and difficulty in focussing.

\textit{Motion parallax} -- is an important depth-cue and relates to the difference in perceived motion between objects placed at a different distance interval from one another~\cite{Woldegiorgis2019,Teng2023}. Incorrect motion parallax can result in disorientation, incorrect perception of depth or self-motion, and incorrect 3D shape perception.

\textit{Field of view (FOV)} -- refers to the observable visual information given a person's point of vision. Monocular human vision, or the field of vision per human eye, equals to 135 degrees vertical and 180 degrees horizontal. For biocular vision, or the combined monocular vision of both eyes, this averages to 210 degrees azimuth approximate, wherein 114 degrees of  overlapping monocular imagery allows for stereoscopic vision and thus depth perception~\cite{Howard2008Jan}. As HMDs sometimes have a display FOV less or different than normal human-vision, depth perception issues can arise~\cite{Willemsen2009Mar,Pfeil2021May}. For instance, a smaller FOV leads to a larger underestimation of depth.

\textit{Latency} -- in displaying visual information within HMDs can arise from hard- and software reasons. Larger latency, ranging from 63 to 101ms, can result in the onset of cybersickness symptoms, in task performance decrease, and a reduction in the sense of body ownership~\cite{Caserman2019Nov}.

\textit{Limitations in color representation} -- can result in e.g., incorrect estimation of relative depth-perception. An example is the inability to discern multiple similarly colored objects placed at different distances from one another when overlapped~\cite{Ping2020Nov}.

\textit{Focal rivalry} -- is mainly an OST issue and arises when digitally rendered content is solely observable on one plane positioned on a different distance from the real-world information observed. This introduces issues with imagery focus of digitally overlayed environmental information~\cite{Condino2019May,Teng2023}.

\textit{Insufficient head tracking accuracy} -- can lead to unsuitability for self-motion use in VEs as an accurate representation of a user's pose with respect to both the physical and virtual world therein is necessary~\cite{Niehorster2017May}.

\textit{Incorrect inter-pupillary distance (IPD)} -- can result in incorrect depth estimation, since stereoscopic vision is affected. A smaller IPD can lead to an overestimation of depth in comparison with a larger IPD~\cite{Woldegiorgis2019,Arefin2022Jun}.

\textit{Display quality} -- does not affect depth judgement when an increase in quality is implemented~\cite{Thompson2004}. However, by lowering the display quality or image resolution within the FOV's periphery, a cybersickness mitigating effect can be obtained due to limiting the overal optical flow experienced within dynamic VEs~\cite{Wu2022Nov}.

\textit{Content} -- refers to the method and exposure of virtual information to a human user. When interacting with either a virtual or real world environment, depth estimation is considerably improved during longer exposure times while using a HMD or conventional stereoscopic display monitors~\cite{Ziemer2009Jul,Naceri2009,Gagnon2021May}.

\textit{Jitter} -- can occur due to a discrepancy between digital rendering and visualization pipelines, and spatial localization in VEs. This can possibly result in eye-strain and fatigue. Perception of jitter artifacts appears to increase with viewing images at larger distances, and can decrease with increased background luminance in VEs~\cite{Wilmott2022}.

\textit{Image registration accuracy} -- is important as misregistration of virtual objects can break visual illusion in MR, lowering immersiveness~\cite{Nabiyouni2017Jan,Wang2017Mar}.

\textit{Exposure and lighting} -- with higher lux than indoor work environments, can lower the visual perception of digitally rendered content in OST HMDs. To counteract this, tinted visors are used within such devices. This in turn introduces a decrease in visibly discernible contrast, either color or luminance based. I.e., incorrect exposure can result in worse contrast perception and lessen discernibility of virtual elements, possibly lowering depth perception in the process~\cite{Erickson2020Oct}.
 
\textit{The absence of shadows} -- can result in underestimation of perceived depth of virtually rendered objects, more so for VST devices~\cite{Adams2022}.

\textit{Lens distortions} -- can result in warped imagery and unintended optical flow, also referred to as dynamic distortion, possibly leading to the onset of cybersickness symptoms~\cite{Chan2022,Xia2023Jan}.

\textit{Lens-eye and camera-eye placement} -- are important, considering that both vertical and horizontal binocular disparity are utilized by the human to infer depth cues. From~\cite{Vienne2016Dec} we find that humans use horizontal disparity for judgment of distance and depth when able to compare objects with background surfaces. If vertical disparity is closer than vergence, distances are understimated. When vertical disparity is further than vergence, distances are overestimated. Thus, the use of close-proximity displays HMDs can result in underestimation of depth due to vertical disparity and vergence not being congruent to one another when observing objects placed at further distances in unclutered environments~\cite{Vienne2020Feb}.

Summarizing the deviations in Table~\ref{tab1_deviations}, it can be concluded that both VST and OST suffer from a wide range of artefacts. However, it should be noted that for OST, the listed deviations are more applicable to the virtual content, instead of the real-world. For VST, as both the real-world aspects and the virtual content are displayed on the same enclosed display, the deviations are applicable to both. Therefore, it can be suggested that real-world interaction is possibly more affected by perceptual deviations using a VST HMD, compared to an OST HMD\@.

\begin{table}[hbt!]
    \footnotesize
    \begin{center}
    \begin{tabular}{lll}
    \toprule
    \textbf{Deviation}                                              & \textbf{Applicable To}    & \textbf{Reference} \\ \midrule
    Vergence-accomodation conflict                                  & VST, OST                  &~\cite{Kramida2016Jul,Zhou2021Dec} \\ 
    Motion parallax                                                        & VST, OST                  &~\cite{Teng2023} \\ 
    Field of view                                                   & VST, OST                  &~\cite{Willemsen2009Mar,Pfeil2021May}\\ 
    Latency                                                         & VST, OST                  &~\cite{Caserman2019Nov} \\ 
    Color                                                           & VST, OST                  &~\cite{Ping2020Nov} \\ 
    Focal rivalry                                                   & OST                       &~\cite{Condino2019May,Condino2022}\\ 
    Head tracking accuracy                                        & VST, OST                  &~\cite{Niehorster2017May}\\ 
    Interpupillary distance                                         & VST, OST                  &~\cite{Woldegiorgis2019,Arefin2022Jun}\\ 
    Display quality                                                 & VST, OST                  &~\cite{Thompson2004} \\ 
    Content                                                         & VST, OST                  &~\cite{Ziemer2009Jul,Naceri2009,Gagnon2021May}\\ 
    Jitter                                                          & VST, OST                  &~\cite{Wilmott2022}\\ 
    Image registration accuracy                                     & VST                       &~\cite{Nabiyouni2017Jan,Wang2017Mar}\\ 
    Latency                                                         & VST, OST                  &~\cite{Nabiyouni2017Jan}\\ 
    Exposure and lighting                                           & VST, OST                  &~\cite{Erickson2020Oct}\\ 
    Shadows                                                         & VST, OST                  &~\cite{Adams2022}\\ 
    Lens distortions                                                & VST, OST                  &~\cite{Chan2022,Xia2023Jan}\\ 
    Lens, camera, display placement                                 & VST, OST                  &~\cite{Vienne2016Dec,Vienne2020Feb}\\ \bottomrule
    \end{tabular}
    \end{center}
    \caption{Visual deviations as found within VST and OST HMDs.}\label{tab1_deviations}
\end{table}

\vspace{-1em}
\subsection{Perceptual deviations of the Varjo XR-3}
\begin{figure}[hbt!]
	\centering
	\includegraphics[width=\linewidth]{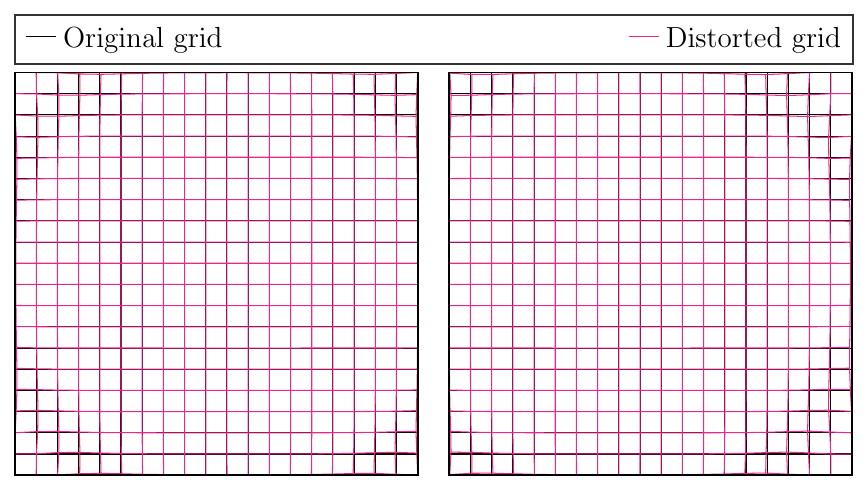}
	\caption{Distortion grid for the VST cameras of the Varjo XR-3.}\label{fig:varjo_camera_distortion}
\end{figure}

\begin{figure}
	\centering
	\begin{subfigure}[t]{0.49\linewidth}
		\centering
		\includegraphics[width=\textwidth]{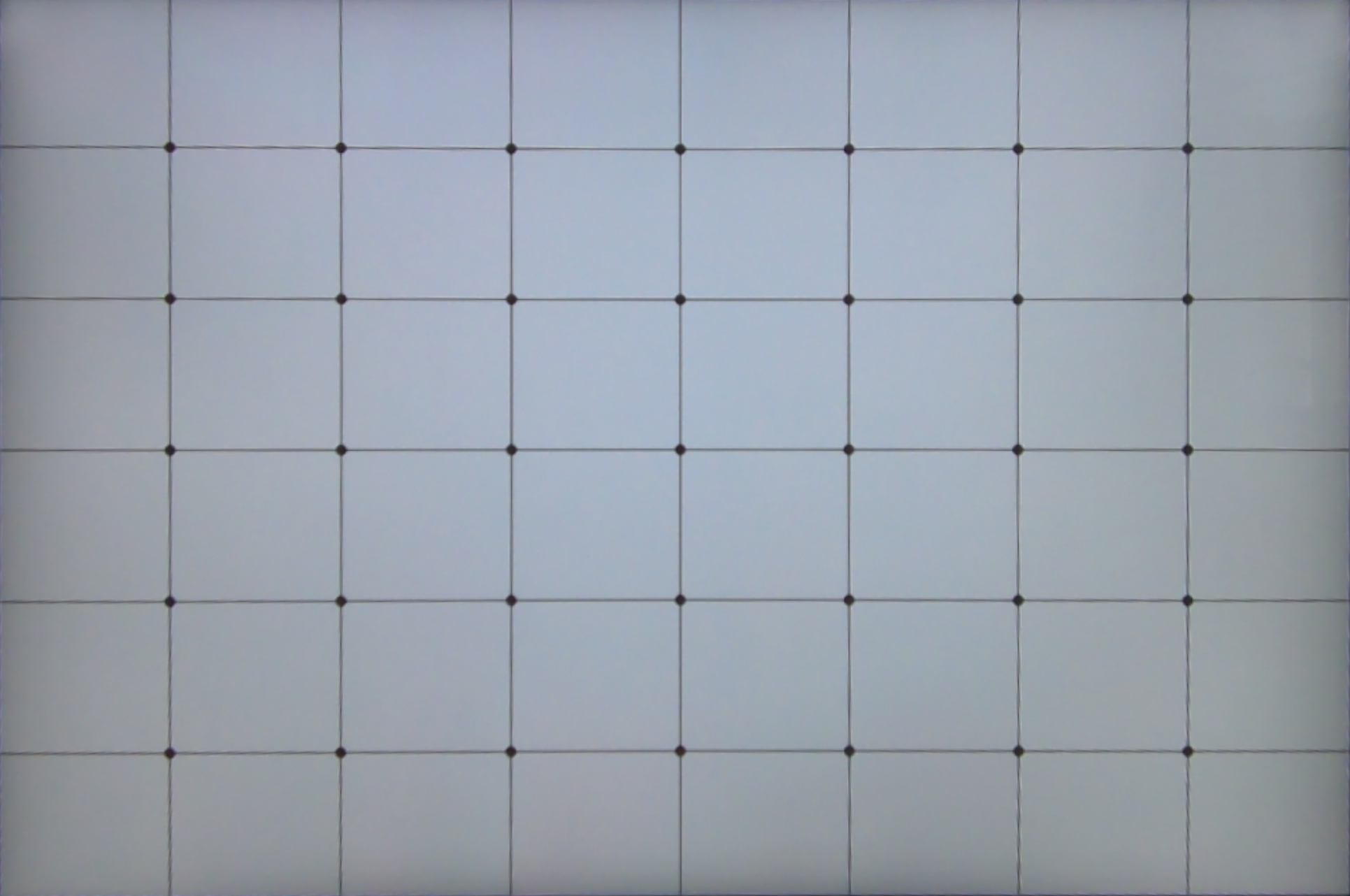}
		\caption{Camera feed from Varjo Base.}\label{fig:varjo_camera_feed}
	\end{subfigure}
	\hfill
	\begin{subfigure}[t]{0.49\linewidth}
		\centering
		\includegraphics[width=\textwidth]{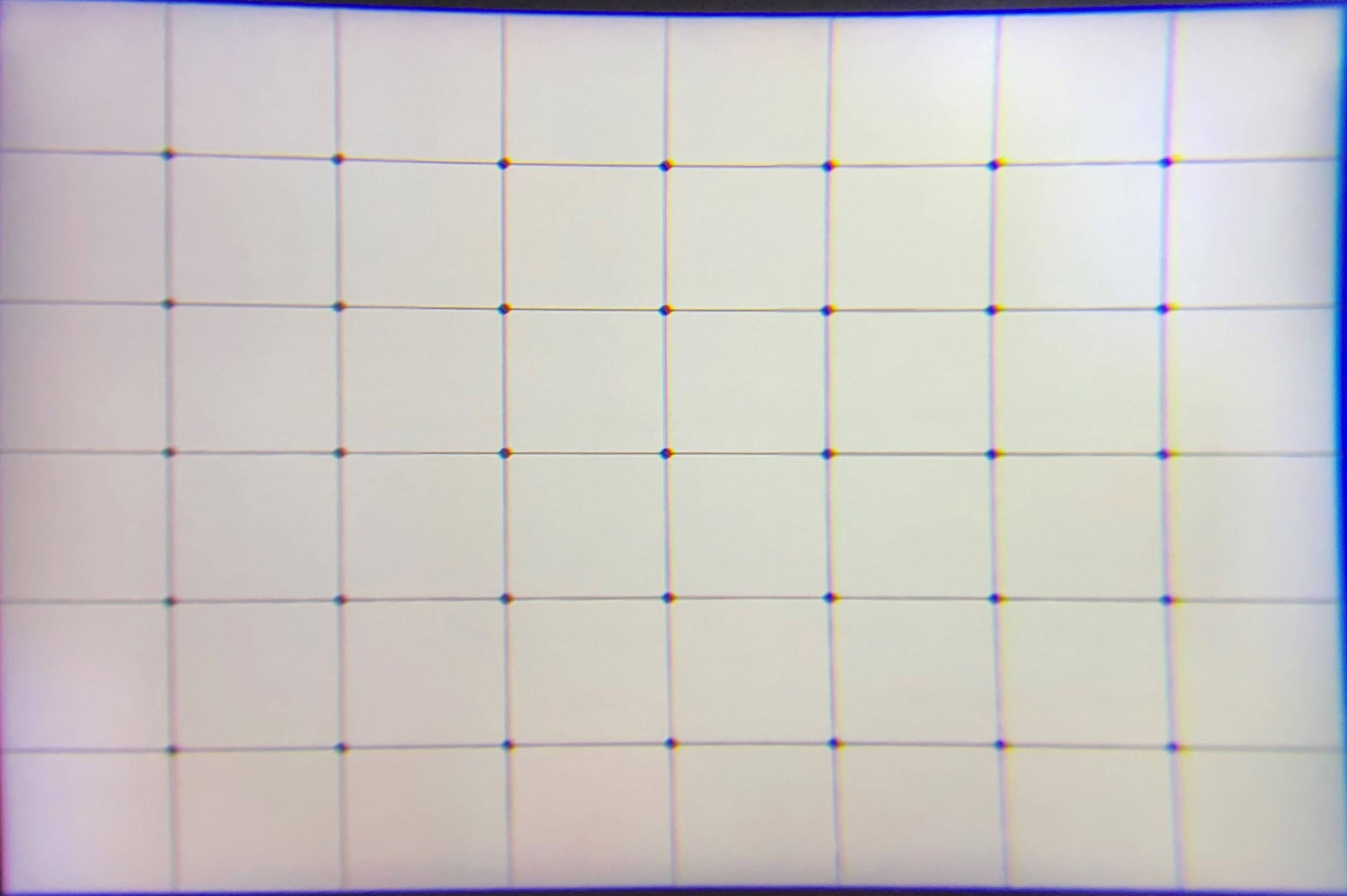}
		\caption{Image through the right eyelens.}\label{fig:varjo_right_lens}
	\end{subfigure}
	   \caption{Varjo XR-3 camera, display, and lens distortions, visualized by displaying a grid on a Microsoft Surface.}\label{fig:varjo_deviations}
\end{figure}

Analyzing the Varjo XR-3's VST design, multiple factors are identified that can introduce perceptual deviations as discussed in Section~\ref{subsub_perceptual_deviations_in_xr}. First, the VST camera positions or the camera extrensics, positioned at $\pm$10~cm in front of the eyes, capture the real world from a different point of vision compared to normal eyesight. In static set-ups, this only results in a slight zooming effect. However, during HMD movements, parallax effects are introduced, as discussed earlier in Section~\ref{subsub_perceptual_deviations_in_xr}. Secondly, the camera intrinsics, including the focal length, aperture, FOV, and resolution, among others, will affect how the cameras translate the real world into pixels. Figure~\ref{fig:varjo_camera_distortion} shows the camera distortion grids of a Varjo XR-3, based on the camera intrinsics retrieved from the Varjo software development kit. Hints of a pincushion distortion are visible, warping the camera imagery slightly around the peripheries.
 When looking at a grid displayed on a Microsoft Surface Studio through the video stream, as shown in Figure~\ref{fig:varjo_camera_feed}, this pincushion distortion appears to be minimal. This videostream is then displayed on two displays per eye, dubbed the Bionic Display\textsuperscript{\texttrademark}, which should mimic human eye-resolution. When observing the same grid through the Varjo's individual ultra-wide non-Fresnel lenses, a clear pincushion distortion is observed, as shown in Figure~\ref{fig:varjo_right_lens}. Moreover, contrast, color offset, and brightness differences are noticed.
Summarizing, deviations perceived while using the Varjo HMD include a pincushion distortion of the peripheries, mainly induced by a combination of the ultra-wide non-Fresnel lenses and Bionic Display\textsuperscript{\texttrademark}, instead of the VST camera intrinsics. However, the exact effect of the camera, lens, and display deviations on the perceived distortion, can not be readily distinguished. Additionally, compared to reality, the real world experienced through the VST HMD is pixelated, includes color offset, has contrast differences, and differs in brightness and intensity.

\section{Experiment}\label{sec:experiment}
To research the identified visual deviations experienced through VST HMDs, as discussed in Section~\ref{sec:background}, a human-in-the-loop experiment was conducted to test a generalizable user-centered measurement setup.

\subsection{Apparatus}\label{sub:experiment_setup}
\begin{figure}[hbt!]
	\centering
	\includegraphics[width=\linewidth]{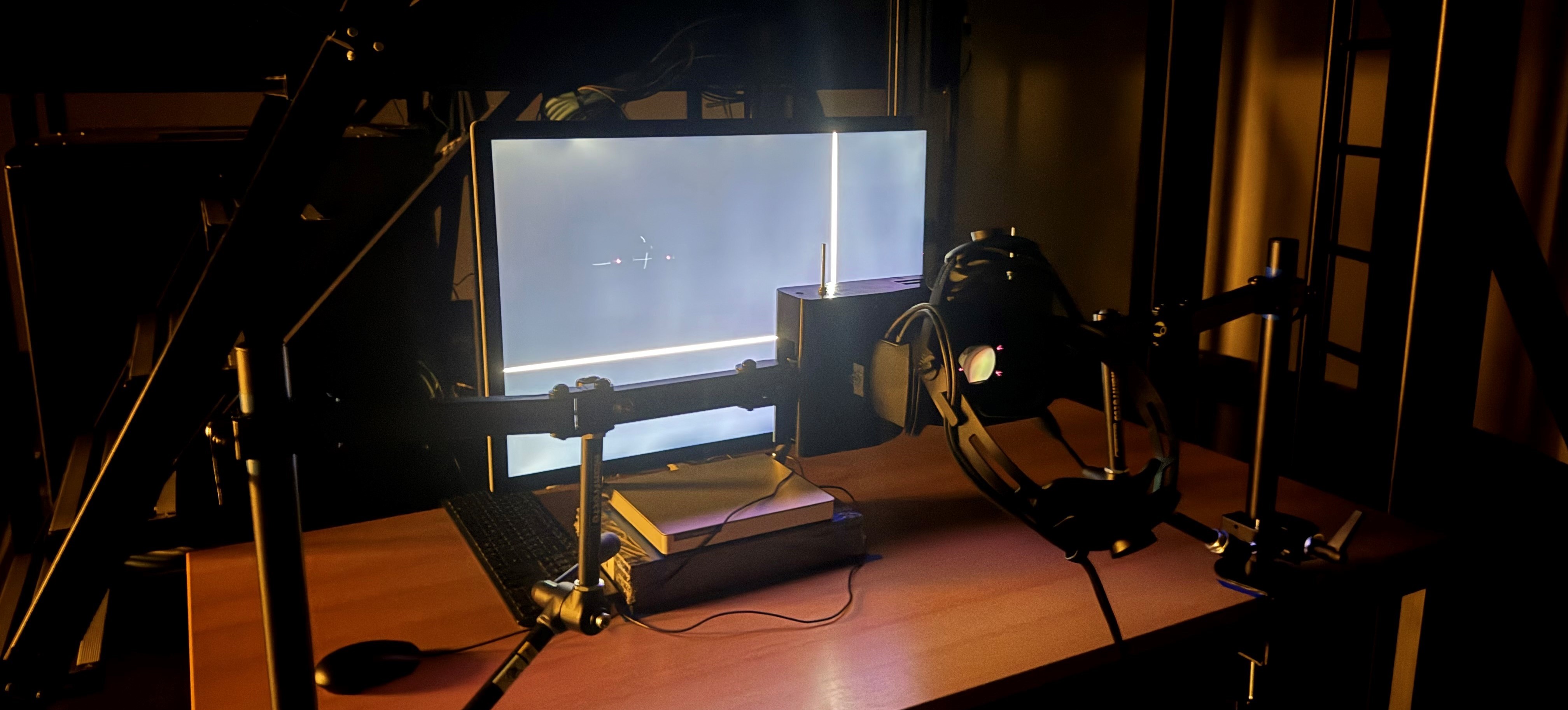}
	\caption{Set-up with a Varjo XR-3 mounted on a static stand.}\label{fig:set_up_dark}
\end{figure}

The experiment setup consisted of a 2018 Microsoft Surface Studio 2 touchscreen desktop computer placed at a distance of 35~cm from a statically mounted Varjo XR-3 VST HMD to ensure that all the display boundaries from the computer could be observed through the HMD\@. This distance also ensured that all participants were able to reach the far boundaries of the touchscreen with both hands, not requiring to adjust their seating position mid-experiment. A custom 3D-printed encasing was used that could be easily attached on and detached from two monitor arms enforced with an additional adjustable boom to ensure rigidity per arm, seen in Figure~\ref{fig:set_up_dark}. The mount was positioned at a height of 33~cm from the tabletop to ensure the HMDs cameras were positioned exactly in the middle of the computer display at 22~cm height and to allow for enough space for all participants to be seated comfortably enough after individual adjustments of stool height.

When no HMD was used during the experiment, a high-end commercial tripod was used as a chin-rest with additional padding to increase comfortability. The U-shape of the tripod mount ensured participants to be wedged ridgidly enough to guarantee no horizontal pivoting head-movements during task execution. Both the HMD mount, as well as the tripod were positioned to be exactly level.

The desktop connected to the Varjo included an Intel Core i9-10940X CPU @ 3.30GHz, NVIDIA RTX A6000, and 64GB RAM, resulting in an application FPS of 89.97 FPS $\pm$ 0.63\% and a display latency of 21.13 ms $\pm$ 3.83\%. The Varjo Base software was 3.5.0.5, the XR-3's firmware version was 0.17.0.469, and the MR cameras firmware version was 48.39.

\vspace{-2em}
\subsection{Participants}\label{sub:participants}
All participants were drafted from a pool of in total 144 potential candidates, all employees or interns at the Royal Netherlands Aerospace Centre (NLR) in Amsterdam. A boundary was set between the age of 18 and 30; the average accepted age for Dutch flight school. In total, 24 participants participated in the study. The age of the participants ranged from 22 to 29 with a mean age of 26.0 years. The participants had an 1:2 female-male ratio, i.e., 8 females and 16 males participated in total. All participants either had normal eyesight or had corrective eye wear assuming normal eyesight, being 4 people with contact lenses and 4 participants with glasses. 2 participants were left-hand dominant while the other 22 participants were right-hand dominant. All gathered data has been used within our data analysis, and no participant data has been omitted.

\subsection{Independent variables and conditions}
The experiment tested one within-participant independent variable, being the type of sight. Two different types of sight were compared in the experiment: (1) the baseline normal eyesight (with contact lenses or glasses if applicable) and (2) the VST-sight utilizing the Varjo XR-3. This resulted in two conditions being the \textit{no-HMD} condition for normal eyesight and the \textit{HMD} condition for the VST-sight while wearing the Varjo XR-3 HMD\@.

\subsection{Task description}\label{sub:task_description}
Defined by evenly partitioning the display, dividing the vertical dimension of the screen into 6 compartments and the horizontal into 8, a sequence of 35 markers were flashed consecutively for 2 seconds each, with a time interval of 3 seconds after disappearance. Participants were instructed to use the 3-second time window to touch the, now black, touchscreen at the position where the marker was perceived to be located prior to dissapearance. Effectively a blind-reaching task, this task repeated itself for 2 sequences per condition, resulting in a total of 70 data points gathered per condition. Only the first touchscreen interaction per marker was recorded. In the case that participants were unable to interact within the 3 second timeframe, the sequence would continue at the pre-defined pace, logging a timeout for the interaction.

\subsection{Dependent measures}\label{sub:dependent_measures}
\begin{figure}[hbt!]
	\centering
	\includegraphics[width=\linewidth]{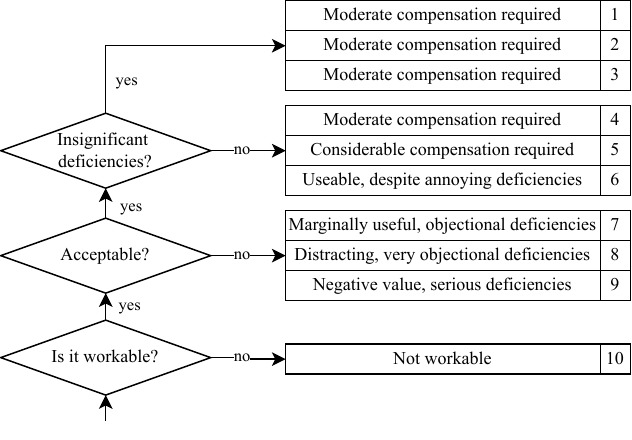}
	\caption{Controllability decision tree inspired on~\cite{Cooper1969pilothandling}\cite{Hodge2015OptimisingPlatform}}\label{fig:controllability_diagram}
\end{figure}

During the experiment, the position of touch interaction was collected expressed in pixels. Secondly, the time between the start of a marker becoming visible and interaction after dissapearance was collected. Moreover, subjective measures were collected. The Simulator Sickness Questionnaire (SSQ)~\cite{Kennedy1993SSQ} was conducted before the start of the experiment and after all experiment conditions. After each condition a MIscery SCale (MISC)~\cite{Bos2006MotionView}, a performance rating (1, \textit{bad} -- 5, \textit{perfect}), and the degree of control adaptation (1, \textit{not having to adapt} -- 5, \textit{certainly having to adapt}) were requested from every participant involved. Finally, for both conditions a controllability rating was asked via a decision tree seen in Figure~\ref{fig:controllability_diagram}, which was constructed based on the Cooper-Harper handling quality scale~\cite{Cooper1969pilothandling} and the Motion Fidelity Rating~\cite{Hodge2015OptimisingPlatform}. Finally, general remarks were asked from the participants.

\subsection{Hypotheses}\label{sub:hypotheses}
Three hypotheses were formulated: 
\begin{enumerate}
	\item[H1.] The overall error between marker and interaction position will be larger for the \textit{HMD} condition as opposed to the \textit{no-HMD} condition.
	\item[H2.] The overall time between marker and interaction position will be larger for the \textit{HMD} condition as opposed to the \textit{no-HMD} condition.
	\item[H3.] Overal interaction accuracy will be increasingly worse for the marker positions placed further into the periphery, more so for with a HMD, but applicable to both conditions.
\end{enumerate}

\subsection{Procedure}\label{sub:procedure}
Prior to the experiment, a written form of consent was gathered from all participants. Participants were notified that the obtained data was kept anynomous and that they have the right to step out of the experiment at any given time. The experiment and task description, apart from the research objective and hypotheses, were all explained to the participants in detail. For the \textit{no-HMD} condition, the chin-rest/tripod was measured to be at the pre-defined height discussed earlier in Section~\ref{sub:experiment_setup}. Participants were instructed to not move their heads, although movements were mostly constrained, and were informed to utilize both hands for touchscreen interaction. The hint was given to use the right hand for markers shown on the right half of the screen, and the left hand for the left half of the screen, in order to allow comfortable interaction therewith. First, familiarization was done without HMD in an illuminated hall. During familiarization, a single experiment run was conducted for the \textit{no-HMD} condition with the markers shown in a sequential order, held consistent over all participants. The actual experiment runs were performed in close to complete darkness as no natural lighting was able to get into the hall, except for a distant lit emergency exit sign. The measurement runs consisted of 2 pseudo-randomized sequences for each condition, i.e., 4 measurement runs per participant. Both the condition order and the sequence order were alternated between participants, resulting in a balanced experiment for all 24 participants. After finishing the entire experiment, participants were informed regarding any questions they might have had and received a small token of appreciation.


\section{Results}\label{sec:results}
The number of missed markers, i.e, the number of unsuccesful touchscreen interactions during a condition, for the \textit{no-HMD} condition ($M = 0.833, SD = 1.204$) compared to the \textit{HMD} condition ($M = 1.125, SD = 1.393)$ resulted in insignificant paired t-test results, $t(24) = -0.942, p = 0.356$. This substantiates the assumption to not include the differences between the number of missed markers into the calculation of the mean errors and response times presented in this section. The missed markers are thus completely omitted from the results.

\begin{figure}[hbt!]
	\centering
	\includegraphics[width=\linewidth]{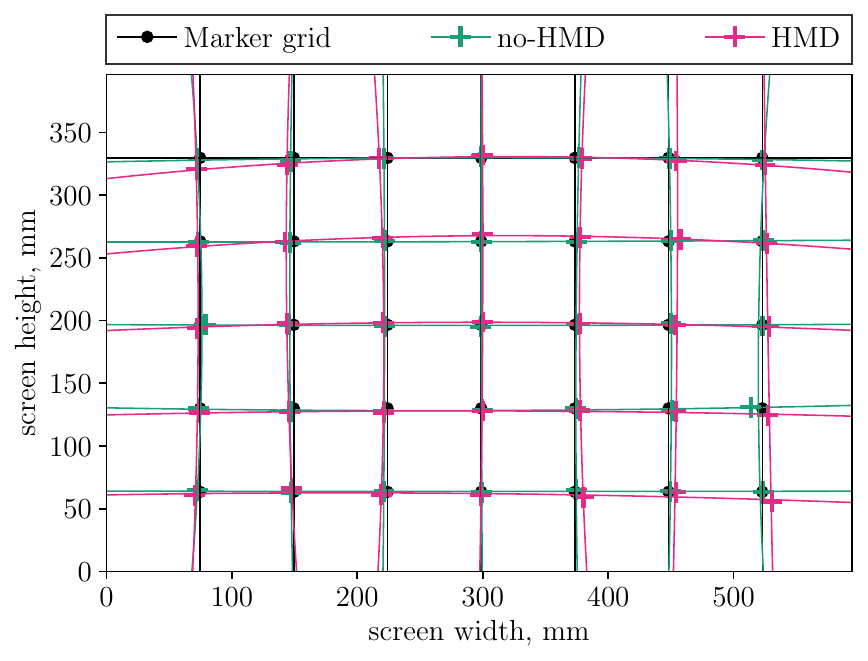}
	\caption{Average interaction location for all participants for both conditions. Grid lines represent a second-order least-squares fit.}\label{fig:average_location_per_marker}
\end{figure}

Figure~\ref{fig:average_location_per_marker} shows the average interaction location per marker based on all participants, where the grid lines represent a second order ploynomial least squares fit. A clear difference between both conditions is visible, where a larger deviation from the original marker positions is seen for the \textit{HMD} condition. Moreover, for the \textit{HMD} condition, larger deviations are seen for the peripheral sections, where the interaction location height is consistently underestimated by the participants.

\begin{figure}[hbt!]
	\centering
	\includegraphics[width=\linewidth]{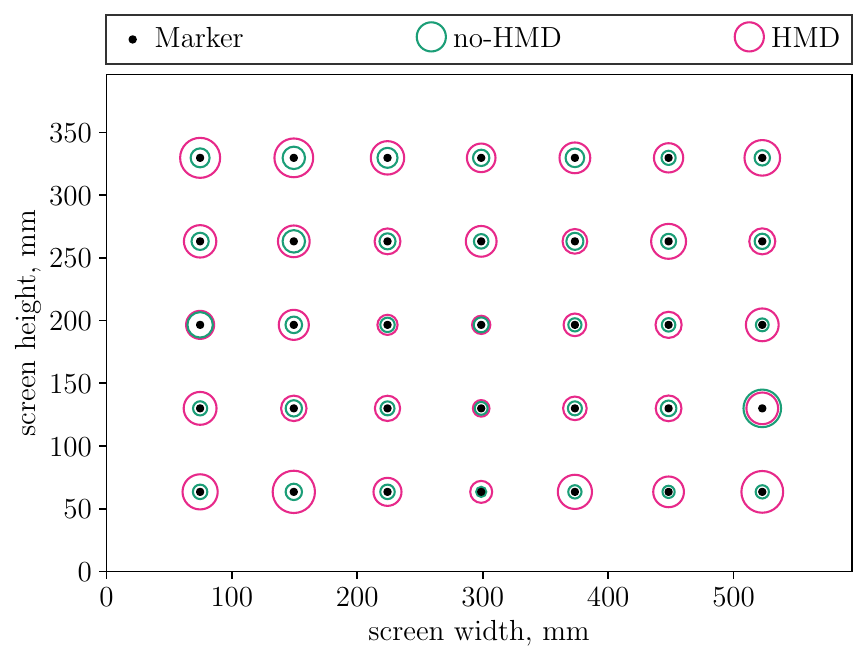}
	\caption{Mean distance error, represented by the circle radius.}\label{fig:average_error_per_marker}
\end{figure}

Figure~\ref{fig:average_error_per_marker} shows the mean distance error for all markers based on all participants, where the radius of the circles represent the mean error in millimeters. Larger errors are found for the \textit{HMD} condition. Moreover, an increasing error size is seen when moving from the center of the screen to the vertical and horizontal edges, which is more apparent for the \textit{HMD} condition compared to the baseline \textit{no-HMD} condition. 

\subsection{Mean distance error and response time}\label{sub:}
\begin{figure}[hbt!]
	\centering
	\begin{subfigure}[t]{0.49\linewidth}
		\centering
		\includegraphics[width=\textwidth]{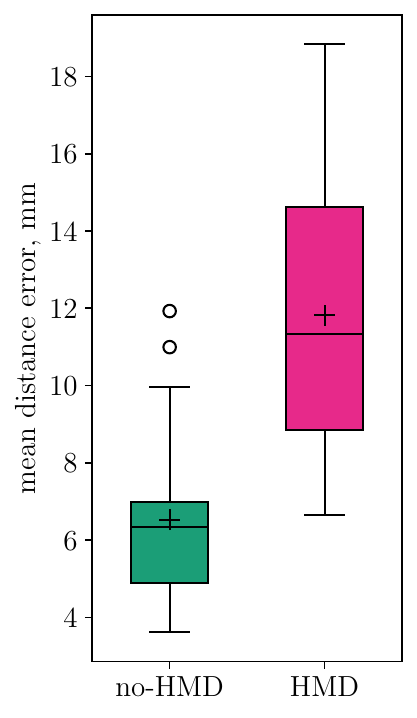}
		\caption{Mean distance error.}\label{fig:box_s_condition}
	\end{subfigure}
	\hfill
	\begin{subfigure}[t]{0.49\linewidth}
		\centering
		\includegraphics[width=\textwidth]{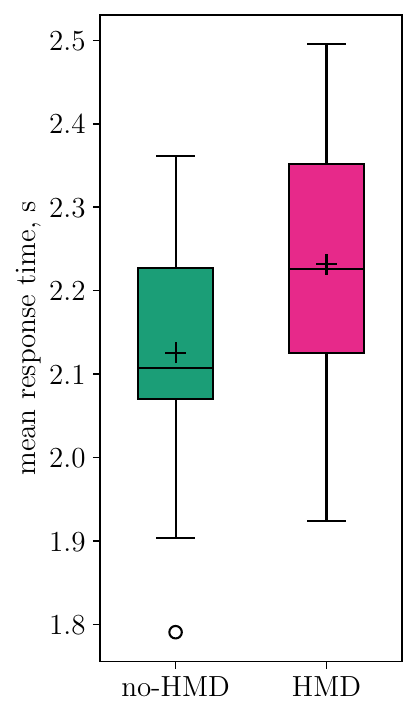}
		\caption{Mean response time.}\label{fig:box_t_condition}
	\end{subfigure}
	   \caption{Box plots including the mean distance error and the mean response time based on all markers for all participants.}\label{fig:box_condition}
\end{figure}

Figure~\ref{fig:box_condition} shows box plots derived from the mean distance error and response time of all markers for all participants. The mean distance error, seen in Figure~\ref{fig:box_s_condition}, shows a 81\% larger mean error for the \textit{HMD} condition. A paired t-test between the mean distance error of the \textit{no-HMD} ($M = 6.529~\si{mm}, SD = 2.131~\si{mm}$) compared to the \textit{HMD} ($M = 11.817~\si{mm}, SD = 3.813~\si{mm}$) indicates a significant difference, $t(24) = -8.762, p = <0.001$. The mean response time, seen in Figure~\ref{fig:box_t_condition}, shows a 5\% longer response time for the \textit{HMD} condition ($M = 2.232~\si{mm}, SD = 0.169~\si{mm}$) compared to the \textit{no-HMD} condition ($M = 2.126~\si{mm}, SD = 0.134~\si{mm}$), resulting in a significant paired t-test result, $t(24) = -3.715, p = <0.001$.

\subsubsection{Intra-screen differences}\label{sub:intrascreen}
Omitting the middle marker-column, a larger distance error is found for the left half of the screen for both conditions, \textit{no-HMD} ($M = 6.994~\si{mm}, SD = 2.144~\si{mm}$) and \textit{HMD} ($M = 12.465~\si{mm}, SD = 4.234~\si{mm}$), compared to the right half of the screen, \textit{no-HMD} ($M = 6.400~\si{mm}, SD = 2.726~\si{mm}$) and \textit{HMD} ($M = 11.985~\si{mm}, SD = 4.439~\si{mm}$). A paired t-test reveiled only significant difference for the \textit{no-HMD} condition, $t(24) = 2.194, p = <0.039$. Similiarly, only significant differences between the left- and right-half response time for the \textit{no-HMD} condition, respectively ($M = 2.150~\si{s}, SD = 0.137~\si{s}$) and ($M = 2.099~\si{s}, SD = 0.139~\si{s}$), is found, $t(24) = 3.588, p = <0.002$. No significant differences are found between the \textit{HMD} left- and right-half response time, respectively ($M = 2.242~\si{s}, SD = 0.167~\si{s}$) and ($M = 2.233~\si{s}, SD = 0.182~\si{s}$).

Omitting the middle marker-row, compared to the bottom-half markers, larger distance error and response times are found for the top-half markers. For the \textit{no-HMD} condition, the distance error for the top-half ($M = 6.932~\si{mm}, SD = 2.085~\si{mm}$) compared to the bottom-half ($M = 6.238~\si{mm}, SD = 3.076~\si{mm}$), is not significant. Likewise, for the \textit{HMD} condition, the distance error for the top-half ($M = 12.592~\si{mm}, SD = 3.849~\si{mm}$) compared to the bottom-half ($M = 11.827~\si{mm}, SD = 5.237~\si{mm}$), is again not significant. For the response time, however, the \textit{no-HMD} condition shows a significant difference $t(24) = 7.831, p = <0.0001$, between the top-half ($M = 2.174~\si{s}, SD = 0.143~\si{s}$) and bottom-half ($M = 2.077~\si{s}, SD = 0.135~\si{s}$). The \textit{HMD} condition shows no significant difference in response time between the top-half and bottom-half, ($M = 2.238~\si{s}, SD = 0.174~\si{s}$) and ($M = 2.244~\si{s}, SD = 0.172~\si{s}$), respectively.

\begin{figure}[hbt!]
	\centering
	\includegraphics[width=\linewidth]{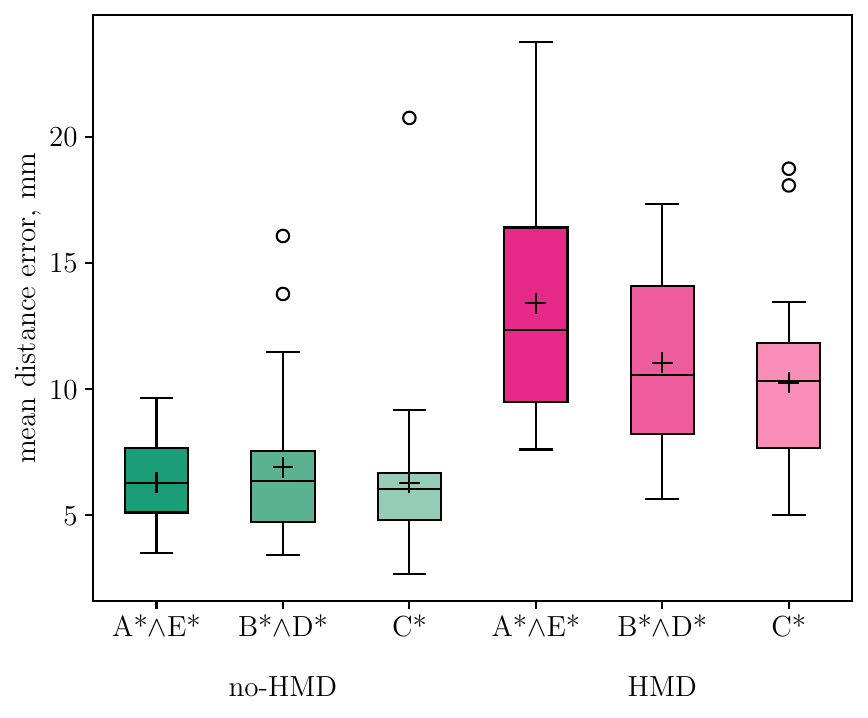}
	\caption{Box plots including the mean distance error based on a specific row of markers for all participants. $A^{*}$ indicates the top-row markers, and subsequent letters indicate descending marker-rows, ending at the bottom-row, marker-row $E^{*}$.}\label{fig:box_s_rows}
\end{figure}

\begin{figure}[hbt!]
	\centering
	\includegraphics[width=\linewidth]{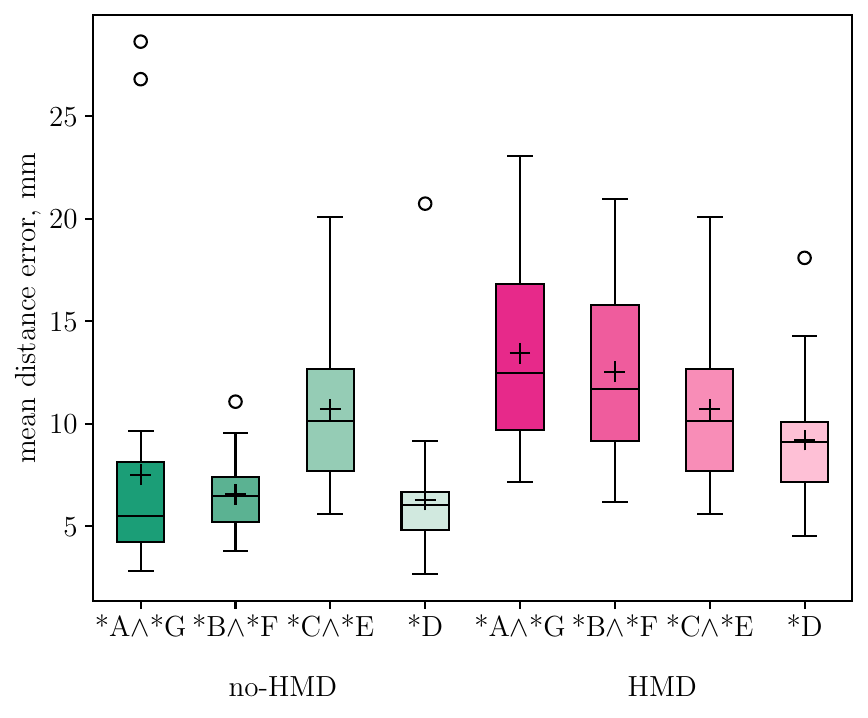}
	\caption{Box plots including the mean distance error based on a specific column of markers for all participants. $^{*}A$ indicates the left-column markers, and subsequent letters indicate increasing right marker-columns, ending at marker-column $^{*}G$, the all-right row markers.}\label{fig:box_s_columns}
\end{figure}

Vertical-wise differences between the mean distance errors are shown in Figure~\ref{fig:box_s_rows}. For the \textit{HMD} condition, the largest distance errors are found near the screen edges, i.e., the top- and bottom-row markers, $A^{*}$ and $E^{*}$, respectively. The distance error for both conditions for rows $A^{*}$ and $E^{*}$ are, ($M = 6.275~\si{mm}, SD = 1.515~\si{mm}$) for the \textit{no-HMD} condition and ($M = 13.394~\si{mm}, SD = 4.813~\si{mm}$) for the \textit{HMD} condition. Subsequently, compared to the top- and bottom-rows, the distance error for the markers on row $B^{*}$ and $D^{*}$ are slightly smaller for the \textit{HMD} condition ($M = 11.032~\si{mm}, SD = 3.823~\si{mm}$), and larger for the \textit{no-HMD} condition ($M = 6.889~\si{mm}, SD = 3.153~\si{mm}$). The middle row markers, $B^{*}$, shows the lowest distance error for the \textit{HMD} condition ($M = 10.242~\si{mm}, SD = 3.466~\si{mm}$), and for the \textit{no-HMD} condition an error of ($M = 6.282~\si{mm}, SD = 3.418~\si{mm}$). One-way repeated measures ANOVA reveals only significant results for the \textit{HMD} condition, $F(2,46) = 13.225, p = <0.001^{*}$ ($^{*}$ indicates corrected for none-sphericity).
Horizontal-wise differences between the mean distance errors are shown in Figure~\ref{fig:box_s_columns}. For the \textit{HMD} condition, the largest distance errors are found near the screen edges ($M = 13.441~\si{mm}, SD = 4.916~\si{mm}$), i.e., the left- and right-column markers, $^{*}A$ and $^{*}G$, respectively. Subsequently, for the \textit{HMD} condition, the distance error decreases when moving more to the middle of the screen, ($M = 12.534~\si{mm}, SD = 4.598~\si{mm}$) for columns $^{*}B$ and $^{*}F$, ($M = 10.705~\si{mm}, SD = 3.868~\si{mm}$) for columns $^{*}C$ and $^{*}E$, and ($M = 9.201~\si{mm}, SD = 2.968~\si{mm}$) for column $^{*}D$. Additionally, an one-way repeated measures ANOVA revealed significant results for the \textit{HMD} condition, $F(3,69) = 14.966, p = <0.001^{*}$ ($^{*}$ indicates corrected for none-sphericity). For the \textit{no-HMD} condition, the largest distance errors are found on columns $^{*}A$ and $^{*}G$ ($M = 7.519~\si{mm}, SD = 6.509~\si{mm}$), followed by columns $^{*}B$ and $^{*}F$ ($M = 6.562~\si{mm}, SD = 1.875~\si{mm}$), ($M = 6.282~\si{mm}, SD = 3.418~\si{mm}$) for middle-column $^{*}D$, and columns $^{*}C$ and $^{*}E$ ($M = 6.176~\si{mm}, SD = 2.143~\si{mm}$). One-way repeated measures ANOVA revealed no significant results for the \textit{no-HMD} condition, $F(3,69) = 0.861, p = < 0.403^{*}$ ($^{*}$ indicates corrected for none-sphericity). 

\vspace{-1em}
\subsubsection{Inter-sequence differences}\label{sub:intersequence}
\begin{figure}[hbt!]
	\centering
	\begin{subfigure}[t]{0.49\linewidth}
		\centering
		\includegraphics[width=\textwidth]{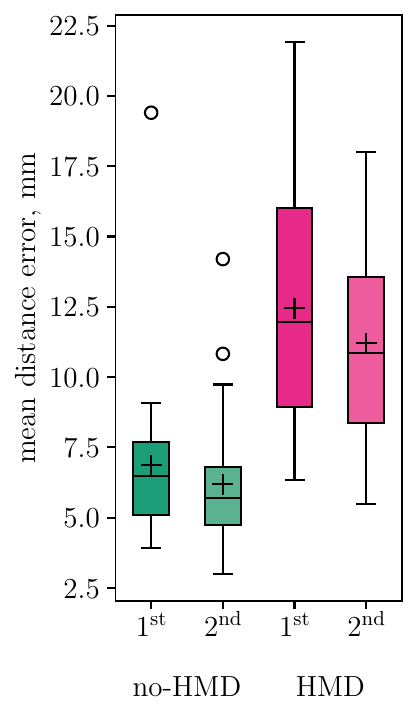}
		\caption{Mean distance error.}\label{fig:box_s_sequence}
	\end{subfigure}
	\hfill
	\begin{subfigure}[t]{0.49\linewidth}
		\centering
		\includegraphics[width=\textwidth]{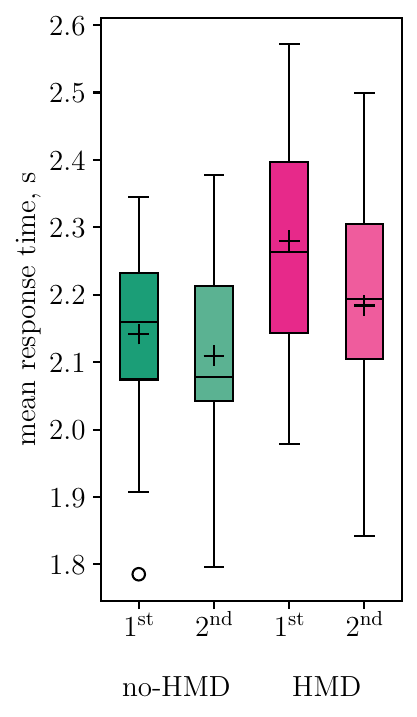}
		\caption{Mean response time.}\label{fig:box_t_sequence}
	\end{subfigure}
	   \caption{Box plots including the mean distance error and the mean response time based on either all first- or second-sequence markers for all participants.}\label{fig:box_sequence}
\end{figure}

Differences between the two sequences for each condition are seen in Figure~\ref{fig:box_sequence}. For both conditions, the distance error and response time decreases for the second sequence. However, paired t-tests revealed no significant difference between the first sequence, ($M = 6.873~\si{mm}, SD = 3.039~\si{mm}$) for the \textit{no-HMD} and ($M = 12.446~\si{mm}, SD = 4.362~\si{mm}$) for the \textit{HMD} condition, and the second sequence, ($M = 6.194~\si{mm}, SD = 2.474~\si{mm}$) and ($M = 11.202~\si{mm}, SD = 3.813~\si{mm}$) for the \textit{no-HMD} and \textit{HMD} condition, respectively. For the response time, significant differences between sequences were found for both conditions. For the \textit{no-HMD} condition, ($t(24) = 2.399, p = 0.025$), with a first sequence of ($M = 2.142~\si{s}, SD = 0.132~\si{s}$) and second sequence of ($M = 2.110~\si{s}, SD = 0.144~\si{s}$). Moreover, for the \textit{HMD} condition, ($t(24) = 4.747, p = <0.001$), with a first sequence of ($M = 2.280~\si{s}, SD = 0.175~\si{s}$) and second sequence of ($M = 2.184~\si{s}, SD = 0.178~\si{s}$).

\subsubsection{Main findings}
In Table~\ref{tab:result_table_s} and Table~\ref{tab:result_table_t}, the mean, standard deviation, and t-test or repeated measures ANOVA results are shown for the distance error and response time respectively. Summarizing, significant distance error and response time differences between conditions are found, where the \textit{no-HMD} reports a smaller error and response time. Significant differences for the markers presented on the left half of the screen compared to the markers on the right half of the screen are only present for the \textit{no-HMD} condition for both the distance error and the response time. Between the top- and bottom half of the screen, only a significant decrease in response time is found for the \textit{no-HMD} condition. Vertical-wise differences are only significant for the \textit{HMD} condition, both the mean distance error and response time are larger towards the edges of the screen. Additionally, for the \textit{HMD} condition, significant horizontal-wise differences, i.e., larger near the edges, are present for the distance error and response time, where for the \textit{no-HMD} condition, only significant differences are found for the response time. To conclude, no significant differences in distance error are reported between the first and second sequence for both conditions. However, the response time is significantly reduced for both conditions for the second sequence.

\begin{table}[hbt!]
	\centering
	\caption{Distance error results, significant results indicated in grey.}\label{tab:result_table_s}
	\footnotesize
	\setlength\extrarowheight{-3pt}
	\begin{tabular}{llrrrr}
	\toprule
							&											&\multicolumn{2}{c}{\textbf{Distance err. (mm)}}&						& 										\\ \cmidrule(lr){3-4}
	\textbf{Condition} 		& \textbf{Test vars} 						& \textbf{Mean} & \textbf{STD} 	 		& \multicolumn{1}{c}{$T$/$F$} 	& \multicolumn{1}{c}{$p$\textbf{-value}}\\
	\midrule
	\multirow{2}{*}{Both} 	& no-HMD 									& 6.529 		& 2.131 		& \multirow{2}{*}{-8.762} 	& \cellcolor{gray!25}								\\ 
							& HMD 										& 11.817 		& 3.813 		&  							& \multirow{-2}{*}{\cellcolor{gray!25}$<$0.001} 	\\\midrule
	\multirow{2}{*}{no-HMD} & left 										& 6.994 		& 2.144 		&  \multirow{2}{*}{ 2.194} 	& \cellcolor{gray!25}  								\\
							& right 									& 6.400 		& 2.726 		&							& \multirow{-2}{*}{\cellcolor{gray!25}0.039}		\\\midrule
	\multirow{2}{*}{HMD} 	& left 										& 12.465 		& 4.234 		&  \multirow{2}{*}{ 0.777} 	& 													\\
							& right 									& 11.985 		& 4.439 		&							& \multirow{-2}{*}{0.445} 							\\\midrule
	\multirow{2}{*}{no-HMD} & top 										& 6.932 		& 2.085 		&  \multirow{2}{*}{ 1.048} 	& 													\\
							& bottom 									& 6.238 		& 3.076 		&							& \multirow{-2}{*}{0.305} 							\\\midrule
	\multirow{2}{*}{HMD} 	& top 										& 12.592 		& 3.849 		&  \multirow{2}{*}{ 0.902} 	&  													\\
							& bottom 									& 11.827 		& 5.237 		&							& \multirow{-2}{*}{0.376}							\\\midrule
	\multirow{3}{*}{no-HMD} & $A^{*}\wedge E^{*}$						& 6.275		 	& 1.515		 	&  \multirow{3}{*}{0.669} 	&  													\\
							& $B^{*}\wedge D^{*}$						& 6.889			& 3.153 		&							& 													\\
							& $C^{*}$									& 6.282			& 3.418 		&							& \multirow{-3}{*}{0.517}							\\\midrule
	\multirow{3}{*}{HMD} 	& $A^{*}\wedge E^{*}$						& 13.394	 	& 4.813		 	&  \multirow{3}{*}{13.225} 	&  	\cellcolor{gray!25}								\\
							& $B^{*}\wedge D^{*}$						& 11.032		& 3.823 		&							& 	\cellcolor{gray!25}								\\
							& $C^{*}$									& 10.242	 	& 3.466 		&							& \multirow{-3}{*}{\cellcolor{gray!25}$<0.001^{*}$}	\\\midrule
	\multirow{4}{*}{no-HMD} & $^{*}A\wedge ^{*}G$						& 7.519		 	& 6.509			&  \multirow{4}{*}{0.861} 	& 													\\
							& $^{*}B\wedge ^{*}F$						& 6.562			& 1.875 		&							& 													\\
							& $^{*}C\wedge ^{*}E$						& 6.176			& 2.143 		&							& 													\\
							& $^{*}D$									& 6.282		 	& 3.418 		&							& \multirow{-4}{*}{$0.403^{*}$}						\\\midrule
	\multirow{4}{*}{HMD} 	& $^{*}A\wedge ^{*}G$						& 13.441		& 4.916 		&  \multirow{4}{*}{14.966} 	&  \cellcolor{gray!25}								\\
							& $^{*}B\wedge ^{*}F$						& 12.534		& 4.598  		&							& 	\cellcolor{gray!25}								\\
							& $^{*}C\wedge ^{*}E$						& 10.705		& 3.868  		&							& 	\cellcolor{gray!25}								\\
							& $^{*}D$									& 9.201		 	& 2.968			&							& \multirow{-4}{*}{\cellcolor{gray!25}$<0.001^{*}$}	\\\midrule
	\multirow{2}{*}{no-HMD} & 1\textsuperscript{st} seq. 				& 6.873 		& 3.039		 	&  \multirow{2}{*}{ 0.947} 	&  													\\
							& 2\textsuperscript{nd} seq. 				& 6.194 		& 2.474 		&							& \multirow{-2}{*}{0.353} 							\\\midrule
	\multirow{2}{*}{HMD} 	& 1\textsuperscript{st} seq. 				& 12.446 		& 4.362 		&  \multirow{2}{*}{ 2.052} 	&  													\\
							& 2\textsuperscript{nd} seq. 				& 11.202 		& 3.813 		&							& \multirow{-2}{*}{0.052} 							\\
	\bottomrule
	\end{tabular}
\end{table}

\begin{table}[hbt!]
	\centering
	\caption{Response time results, significant results indicated in grey.}\label{tab:result_table_t}
	\footnotesize
	\setlength\extrarowheight{-3pt}
	\begin{tabular}{llrrrr}
	\toprule
	&											&\multicolumn{2}{c}{\textbf{Response time (s)}}&						& 										\\\cmidrule(lr){3-4}
	\textbf{Condition} 		& \textbf{Test vars} 	& \textbf{Mean} & \textbf{STD}  & \multicolumn{1}{c}{$T$/$F$} 	& \multicolumn{1}{c}{$p$-value}\\
	\midrule
	\multirow{2}{*}{Both} 	& no-HMD 									& 2.126 		& 0.134 		&  \multirow{2}{*}{-3.715} 	& \cellcolor{gray!25} 								\\
							& HMD										& 2.232 		& 0.169 		&  							& \multirow{-2}{*}{\cellcolor{gray!25}0.001} 		\\\midrule
	\multirow{2}{*}{no-HMD} & left 										& 2.150 		& 0.137 		&  \multirow{2}{*}{ 3.588} 	& \cellcolor{gray!25} 								\\
							& right 									& 2.099 		& 0.139 		&							& \multirow{-2}{*}{\cellcolor{gray!25}0.002} 		\\\midrule
	\multirow{2}{*}{HMD} 	& left 										& 2.242 		& 0.167 		&  \multirow{2}{*}{ 0.564} 	& 													\\
							& right 									& 2.233 		& 0.182 		&							& \multirow{-2}{*}{0.578} 							\\\midrule
	\multirow{2}{*}{no-HMD} & top 										& 2.174 		& 0.143 		&  \multirow{2}{*}{ 7.831} 	& \cellcolor{gray!25}								\\
							& bottom 									& 2.077 		& 0.135 		&							& \multirow{-2}{*}{\cellcolor{gray!25}$<$0.001} 	\\\midrule
	\multirow{2}{*}{HMD} 	& top 										& 2.238 		& 0.174 		&  \multirow{2}{*}{-0.518} 	& 													\\
							& bottom 									& 2.244 		& 0.172 		&							& \multirow{-2}{*}{0.609} 							\\\midrule
	\multirow{3}{*}{no-HMD} & $A^{*}\wedge E^{*}$						& 2.128 		& 0.140			&  \multirow{3}{*}{0.172} 	&  													\\
							& $B^{*}\wedge D^{*}$						& 2.123			& 0.133 		&							& 													\\
							& $C^{*}$									& 2.128		 	& 0.137 		&							& \multirow{-3}{*}{0.842}							\\\midrule
	\multirow{3}{*}{HMD} 	& $A^{*}\wedge E^{*}$						& 2.257		 	& 0.174			&  \multirow{3}{*}{13.443} 	&  	\cellcolor{gray!25}								\\
							& $B^{*}\wedge D^{*}$						& 2.226			& 0.172 		&							& 	\cellcolor{gray!25}								\\
							& $C^{*}$									& 2.193		 	& 0.173 		&							& \multirow{-3}{*}{\cellcolor{gray!25}$<0.001$}		\\\midrule
	\multirow{4}{*}{no-HMD} & $^{*}A\wedge ^{*}G$						& 2.104		 	& 0.140		 	&  \multirow{4}{*}{2.214} 	&  													\\
							& $^{*}B\wedge ^{*}F$						& 2.117			& 0.145 		&							& 													\\
							& $^{*}C\wedge ^{*}E$						& 2.123			& 0.137 		&							& 													\\
							& $^{*}D$									& 2.128		 	& 0.137 		&							& \multirow{-4}{*}{0.094}							\\\midrule
	\multirow{4}{*}{HMD} 	& $^{*}A\wedge ^{*}G$						& 2.269		 	& 0.192		 	&  \multirow{4}{*}{9.837} 	&  \cellcolor{gray!25}								\\
							& $^{*}B\wedge ^{*}F$						& 2.240			& 0.166 		&							& 	\cellcolor{gray!25}								\\
							& $^{*}C\wedge ^{*}E$						& 2.205			& 0.168 		&							& \cellcolor{gray!25}								\\
							& $^{*}D$									& 2.195		 	& 0.175 		&							& \multirow{-4}{*}{\cellcolor{gray!25}$<0.001$}		\\\midrule
	\multirow{2}{*}{no-HMD} & 1\textsuperscript{st} seq. 				& 2.142 		& 0.132 		&  \multirow{2}{*}{ 2.399} 	& \cellcolor{gray!25} 								\\
							& 2\textsuperscript{nd} seq. 				& 2.110 		& 0.144 		&							& \multirow{-2}{*}{\cellcolor{gray!25}0.025} 		\\\midrule
	\multirow{2}{*}{HMD} 	& 1\textsuperscript{st} seq. 				& 2.280 		& 0.175 		&  \multirow{2}{*}{ 4.747} 	& \cellcolor{gray!25} 								\\
							& 2\textsuperscript{nd} seq. 				& 2.184 		& 0.178 		&							& \multirow{-2}{*}{\cellcolor{gray!25}$<$0.001} 	\\
		\bottomrule
	\end{tabular}
\end{table}

\subsection{MISC, performance, controllability, and commentary}\label{sub:subjective_results}
\begin{figure}[hbt!]
	\centering
	\begin{subfigure}[t]{0.49\linewidth}
		\centering
		\includegraphics[width=\textwidth]{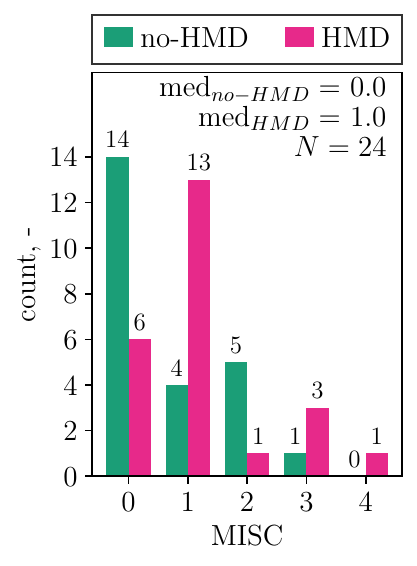}
		\caption{MISC, ranging from 0 to 10, count. Note that no MISC above 4 was reported.}\label{fig:bar_misc}
	\end{subfigure}
	\hfill
	\begin{subfigure}[t]{0.49\linewidth}
		\centering
		\includegraphics[width=\textwidth]{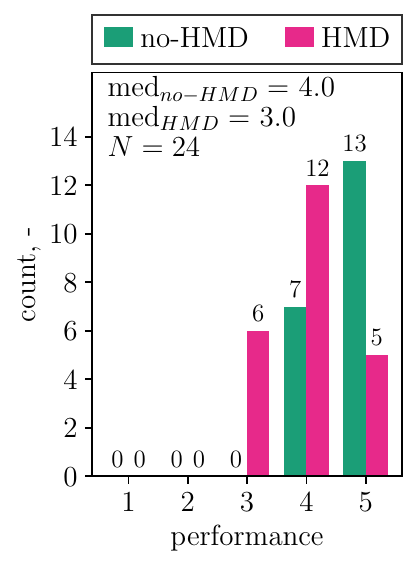}
		\caption{Performance rating count, ranging from 1, \textit{bad}, to 5, \textit{good}.}\label{fig:bar_performance}
	\end{subfigure}
		\caption{MISC and performance ratings.}\label{fig:bar_misc_performance}
\end{figure}

\begin{figure}[hbt!]
	\centering
	\begin{subfigure}[t]{0.49\linewidth}
		\centering
		\includegraphics[width=\textwidth]{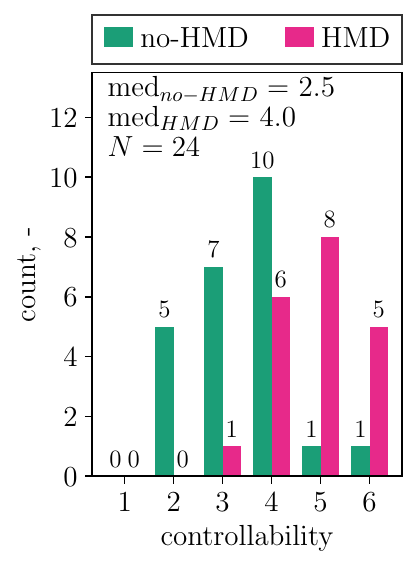}
		\caption{Controllability rating count (ranging 1 to 10, see Subsection~\ref{sub:dependent_measures}).}\label{fig:bar_control}
	\end{subfigure}
	\hfill
	\begin{subfigure}[t]{0.49\linewidth}
		\centering
		\includegraphics[width=\textwidth]{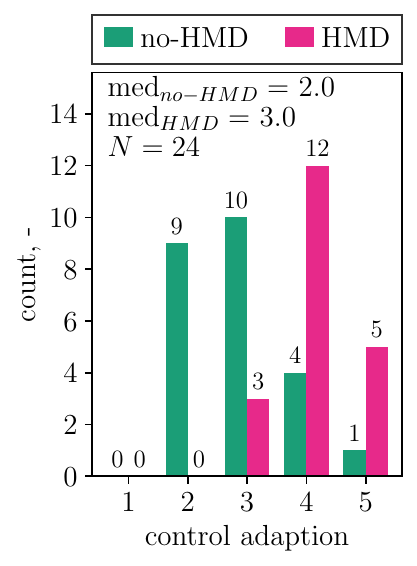}
		\caption{Control adaptation rating count (ranging 1, \textit{not having to adapt}, to 5, \textit{certainly having to adapt}).}\label{fig:bar_adapt}
	\end{subfigure}
		\caption{Controllability and control adaptation ratings.}\label{fig:bar_control_adapt}
\end{figure}

\noindent
Figure~\ref{fig:bar_misc} shows the MISC count distribution for both conditions, where higher ratings for the \textit{HMD} condition are seen. MISC ratings gathered after every condition resulted in a median rating of 0 for the \textit{no-HMD} condition and a median of 1 for the \textit{HMD} condition. A Wilcoxon signed-rank test revealed a statistically significant difference between conditions, ($N=24, Z=14, p=0.008$). The performance rating count distribution, shown in Figure~\ref{fig:bar_performance}, shows higher ratings for the \textit{no-HMD} condition. This is also reflected by the median values of 4 and 3 for the \textit{no-HMD} and \textit{HMD} condition, respectively. A Wilcoxon signed-rank test revealed a statistically significant difference between conditions, ($N=24, Z=8.5, p=<0.001$).
The controllability ratings, seen in Figure~\ref{fig:bar_control}, indicates higher ratings, i.e., less controllable, for the \textit{HMD} condition, with a median of 4, compared to the \textit{no-HMD} condition, with a median of 2.5. A Wilcoxon signed-rank test revealed a statistically significant difference between conditions, ($N=24, Z=0, p=<0.001$). Figure~\ref{fig:bar_adapt} shows the control adaptation rating distribution, 1 to 5, ranging from \textit{not having to adapt} to \textit{certainly having to adapt}. More control adaptation was required for the \textit{HMD} condition with a median of 3, compared to the \textit{no-HMD} condition with a median of 2. Again, a Wilcoxon signed-rank test revealed a statistically significant difference between conditions, ($N=24, Z=0, p=<0.001$).
Overall cybersickness syptoms, collected from the SSQ, were not found before the experiment. However, after the experiment, one-third of the participants reported symptoms such as slight general discomfort, slight eyestrain and slight to severe fullness of head.

From these results we gather that, although only slight cybersickness symptoms were identified in participants, these were experienced more so after the \textit{HMD} condition as opposed to the \textit{no-HMD} condition. Participants vastly underestimated their performance with the use of the HMD, feeling less in control overall versus not utilizing an HMD, and felt more need to adjust their control strategy in approaching the experiment task. From participant comments we found that the overall discomfort contributing to cybersickness was experienced due to being statically constrained either by the chin-rest in the \textit{no-HMD} condition, or when wearing the mounted HMD during the \textit{HMD} condition.

\section{Discussion}\label{sec:discussion}
The experiment hypotheses, presented in Section~\ref{sub:hypotheses}, encompass the distance error between the marker and interaction location, the response time, and the intra-screen distance error differences. Based on the results in Section~\ref{sec:results}, significant proof for all three hypotheses have been given: (1) The overall error between marker and interaction position is larger for the \textit{HMD} condition as opposed to the \textit{no-HMD} condition. (2) The overall response time between marker and interaction moment is longer for the \textit{HMD} condition as opposed to the \textit{no-HMD} condition. (3) Overal interaction accuracy is increasingly worse for the marker positions placed further into the periphery, more so for the \textit{HMD} condition, but applicable to both conditions.

Significant proof for the first hypothesis comes from a combination of observed perceptual deviations. One deviation in the \textit{HMD} condition, differing from normal eyesight, is the image brightness and light intensity, confirmed by a participant commenting on the \textit{HMD} condition being \textit{``easier on the eyes''}. During the \textit{no-HMD} condition, the high-contrast and bright markers displayed on the touchscreen, in combination with the darkened room, caused the markers to still be visible on the retina after disappearance. Thus, opposed to the \textit{HMD} condition, for the \textit{no-HMD} condition, the location of the markers was even visible when participants reached for the marker location on the touchscreen. Secondly, depth deviations caused by e.g., the VST camera extrensics, result in the touchscreen to appear larger in comparison with normal eye-sight. Lastly, the VST camera and the HMD lens distortions warp real-world information to show a pincushion effect. These last two deviations resulted in the markers to be perceived by the participants at a different location than in the real-world.

The second hypothesis outcome is explained by participant comments. Half of the participants indicated to have depth perception issues while wearing the HMD\@. Resulting in underestimation of depth, this caused unconfidence in their interaction movement compared to normal eyesight. Results suggest that longer exposure to the HMD improves response time. However, the same can be argued for task familiarization, since response time improvements were also found for the \textit{no-HMD} condition.

An explanation for the third hypothesis cannot be given exclusively. Looking at the results in Figure~\ref{fig:average_error_per_marker}, it can be seen that the average error per participant is larger at the peripheries of the touchscreen correlating with larger peripheral deformities present in the HMD, seen in Figure~\ref{fig:varjo_right_lens}. It can be argued that the average interaction error is positively correlated with the degree of compounded camera, display, and lens distortion as found within the Varjo. However, the deviation grid in Figure~\ref{fig:average_location_per_marker} is not similar to the pincushion distortion in Figure~\ref{fig:varjo_right_lens}. Thus, although larger errors in the peripheries are found, this does not result into conclusive evidence that either camera, display, or lens distortion is the main culprit for the interaction deviation.

No significant decrease in distance error was found over time, even though the response time did decrease significantly. Experiment task familiarization could have made participants more confident and faster in performing the task. However, the absence of visual feedback, due to the hands and marker not being visible simultaneously, ensured that the participants could not decrease the distance errors over time.

Reflecting back on the experiment design, one can argue that the differences in the perceptual brightness and intensity enlarged the discrepancy between conditions, removing the focus from the perceptual deviations caused by camera and lens distortions. One method to counteract this within future experiments is to interact with the markers while still being visible, ensuring that the markers remain on the retina for both conditions. However, when hands are visible visual feedback thereof can possibly mask the perceptual deviations of interest. Counteracting the visual hand feedback can be done by pre-recording the experiment with the HMD camera stream, and replaying this recording while the participants interact with a physical touchscreen. Somewhat similarly, to solely assess the HMD lens distortions a virtual copy of the experiment set-up alongside its physical counterpart can be created, such that participants only see the virtual replica while interacting with the physical set-up. This way, both the hands and camera feed distortions are not visible. 

Not explicitly studied in this work, future research opportunities in VST HMD depth perception deviations studies arise. These deviations were identified as an important culprit by half of the participants regarding decreased task performance for the \textit{HMD} condition. Perceptual depth deviations could be significant. For example, the zooming effect visual deviation due to the HMD's camera extrensics can have an effect on nearby object perception. Thus, it can be argued that the individual nor compounded effect of visual deviations, such as (camera) lens intrinsics, extrensics, brightness, and depth deviations, on physical object manipulation task performance remains unclear. Additionally, repeating the experiment with another HMD, which should have a considerable different lens distortion, can conclude whether the average interaction locations and resulting grid lines, shown in Figure~\ref{fig:average_location_per_marker}, are results of the VST camera and lens distortions, or other deviations present in an HMD\@. Lastly, this study focussed on measuring perceptual deviations, however did not yet tackle the issue of counteracting those. Thus, based on the results found within this study, counteractive measures in order to mitigate the identified perceptual deviations should be researched. E.g., one could look into HMD-specific visual imagery adjustment, or even on individual level, based on results from the measurement method of this study. Visual imagery adjustment could be done by e.g., attempting to nullify the identified HMD-specific pinchushion distortion via a barrel distortion type warping. Both methods, with and without imagery adjustment, could then be compared to conclude if a closer approximation to normal eyesight vision holds beneficiary potential for real-world interaction or not.

\section{Conclusion}\label{sec:conclusion}

In this paper we proposed a static set-up and human-centered method for evaluating perceptual deviations in VST HMDs. Objective results indicate that the mean error and interaction time per participant when interacting with a physical touchscreen was higher when using a HMD versus wearing none. It was found that for both conditions the participants' interaction time decreases, without the interaction error changing over time. Reasoning suggests habituation but no performance increase due to the absence of haptic or visual feedback. It is suggested that pincushion distortion of the real-world imagery as observed through the Varjo XR-3 HMD could be a reason for the higher average peripheral interaction error, however conclusive evidence needs to be obtained from future studies. Subjective results indicate that the overall sense of performance, comfort, and controllability was less when using a HMD\@. Moreover, it was found that users felt greater need to adjust their control strategy as opposed to wearing no HMD\@.

On the basis of this paper, future work is recommended to extend on the proposed method to either re-perform the exact experiment comparing other VST HMDs to proof the generalizability of our proposed method. Based hereupon, researchers can look into mitigating or counteractive measures, being either hard- or software based solutions to tackle measured deviations. The setup as proposed within this paper, can then be used as a comparative tool to discern whether the to be identified countermeasures prove to be a significant addition or not. However, as our results indicate no clear evidence for the need to correct for visual deviations in VST HMDs when observing real-world information using a static setup, we can not account for these possible static countermeasures to counteract visual deviations in dynamic scenarios as well. This should also be researched.

\acknowledgments{This work was conducted by the NLR and funded by the Dutch MoD. Resulting from a joint-project between the NLR, TNO and MARIN, their feedback regarding the experiment is acknowledged. However, all work done is accredited to the authors, where the first two authors contributed equally.}

\bibliographystyle{abbrv-doi}

\bibliography{bib}
\end{document}